\documentclass[aps,prstab,reprint,superscriptaddress]{revtex4-2}
\usepackage{graphicx}% Include figure files
\usepackage{amsmath}
\usepackage[colorlinks=false,urlcolor=blue]{hyperref}
\usepackage{cleveref}
\usepackage{chngcntr}
\usepackage{float}

\begin{document}

%\preprint{}

\title{Evolution of Transversely Asymmetric Electron Beams in Hollow Plasma Channels}


\author{Siqin Ding}
\affiliation{%
	Department of Engineering Physics, Tsinghua University, Beijing 100084, China}

\author{Shiyu Zhou}%
\email{zhousy@ihep.ac.cn}
\affiliation{%
	Institute of High Energy Physics, Chinese Academy of Sciences, Beijing 100049, China}
    
\author{Fei Li}%
\affiliation{%
	Institute of High Energy Physics, Chinese Academy of Sciences, Beijing 100049, China}

\author{Jianfei Hua}
\affiliation{%
	Department of Engineering Physics, Tsinghua University, Beijing 100084, China}
	
\author{Wei Lu}
\email{weilu@ihep.ac.cn}
\affiliation{%
	Institute of High Energy Physics, Chinese Academy of Sciences, Beijing 100049, China}

\date{\today}

\begin{abstract}
Plasma wakefield acceleration in hollow plasma channels has emerged as a promising approach for positron acceleration, since an electron beam can drive wakes with a transversely uniform accelerating field and no intrinsic defocusing force for positrons. Recently, it was proposed that a transversely asymmetric electron beam can excite quadrupole-dominated wakefield in a hollow channel, enabling the formation of accelerating and focusing fields suitable for positrons [S. Zhou, \textit{et al.}, \href{https://doi.org/10.1103/PhysRevLett.127.174801}{Phys. Rev. Lett. 127, 174801 (2021)}]. However, the self-consistent evolution and stability of such asymmetric drivers, which are crucial for sustaining a usable wake over long distances, remain insufficiently understood. In this work, we investigate the evolution of wakefield driven by asymmetric electron beams in hollow plasma channels using fully three-dimensional particle-in-cell simulations. We identify two distinct unstable scenarios: a reversal of quadrupole field polarity and continuous penetration of the driver into the plasma wall. By analyzing the transverse dynamics of the driver and the restoring forces provided by the channel ions, we establish simple physical criteria that ensure stable propagation. These results clarify the fundamental constraints governing asymmetric-driver evolution, provide practical guidance for realizing long-lived, quasisteady wakes in hollow plasma channels, and indicate that the scheme can preserve mm$\cdot$mrad-level normalized positron emittance.
\end{abstract}

%\keywords{}

\maketitle

\section{Introduction \label{Sec:I}}
The discovery of the Higgs boson has intensified global efforts toward next-generation electron–positron colliders, such as the ILC, CEPC, and FCC-ee \cite{CEPC_2019_Lou,FCC_2019_Benedikt,ILC_2019_Michizono}, aimed at precision Higgs measurements and searches beyond the Standard Model. However, the enormous size and cost associated with radio-frequency (RF) technology suggest that these facilities may represent the practical limit of conventional accelerator concepts. To extend the energy frontier beyond the Higgs scale in a cost-effective manner, alternative acceleration schemes capable of providing much higher gradients are actively being explored. Among them, plasma-based acceleration (PBA) \cite{LWFA_1979_Tajima,PWFA_1985_Chen} has emerged as a promising candidate due to its ultra-high accelerating gradients and rapidly improving beam quality. While PBA in the nonlinear blowout regime \cite{blowout_1991_Rosenzweig,blowout_2006_Lu} has demonstrated multi-GeV acceleration with excellent beam quality for electrons \cite{42GeV_2007_Blumenfeld,10GeV_2024_Picksley,High_charge_2020_Gotzfried,High_efficiency_2014_Litos,High_efficiency_2021_Lindstrom,Low_emittance_2012_Plateau,Low_spread_2021_Ke,Injection_2021_Wu}, extending similar performance to positrons remains a major challenge. In the blowout regime, the focusing and accelerating phases favorable for positrons are intrinsically limited, often leading to strong nonlinear fields, large energy spread, or emittance growth. Meeting the stringent requirements of future linear colliders\cite{ILC_TDR_2013,FCCee_2022,CEPC_TDR_2024}—high charge, low emittance, low energy spread, and high energy-transfer efficiency—therefore demands fundamentally different wakefield structures for positron acceleration\cite{LG_2014_Vieira,DualGaussian_2014_Yu,Hollow_beam_2015_Jain,RingGaussian_2020_Xu,LWFA_beamloading_2023_Liu,Finiteplasma_2019_Diederichs,PWFA_beamloading_Zhou_2025}.

A hollow plasma channel has been proposed as an attractive configuration for positron acceleration, as it can, in principle, provide uniform accelerating force and no defocusing force for positrons \cite{Hollow_channel_1995_Chiou,Hollow_channel_1998_Chiou,Hollow_channel_1999_Schroeder,Hollow_channel_2013_Schroeder,Hollow_channel_2016_Gessner}. The experimental generation of such hollow plasma channels can be practically realized using advanced laser-shaping techniques. Specifically, high-order Bessel beams can be utilized to pre-ionize a well-defined annular region of a neutral gas target \cite{Hollow_channel_generation_1996_Andreev, Hollow_channel_2016_Gessner, Hollow_channel_generation_2000_Fan, Hollow_channel_2024_Liu}. However, the absence of background plasma ions and electrons inside channel eliminates the intrinsic focusing force, rendering both driver and witness beams susceptible to beam breakup instabilities. Recently, a novel scheme has been proposed in which a transversely asymmetric electron beam propagating on axis excites quadrupole-dominated wakefield inside a hollow plasma channel\cite{Hollow_channel_2021_Zhou}. As the driver splits in the quadrupole field and reach the inner boundary of the channel, simultaneous accelerating and focusing fields for positrons emerge. Initial simulations have demonstrated promising positron acceleration with moderate energy spread and controlled emittance growth. Despite these encouraging results, a fundamental and largely unexplored question remains: under what conditions can a transversely asymmetric electron driver self-consistently evolve into a long-lived, quasisteady wake in a hollow plasma channel? The stability of the driver is a prerequisite for any practical positron acceleration scheme, yet the evolution of asymmetric drivers is intrinsically complex. Improper beam parameters may lead to wake distortion which destroys the desired wake structure. To date, the physical origins of these behaviors and the boundaries between stable and unstable regimes have not been systematically investigated.

In this work, we focus on the self-consistent evolution and stability of quadrupole-dominated plasma wake driven by transversely asymmetric electron drivers in hollow plasma channels. Using fully three-dimensional particle-in-cell (3D PIC) simulations, we identify distinct evolution regimes of the driver–wake system and uncover two dominant mechanisms associated with (i) transverse phase-space mismatch of the driver and (ii) excessive transverse kinetic energy relative to the channel’s confinement. By analyzing the transverse dynamics of the driver and the restoring forces provided by the channel ions, we establish simple and physically transparent criteria that are indicative of stable operating regimes. These results establish the fundamental constraints governing asymmetric-driver operation and provide practical guidance for sustaining quasisteady wakefield in hollow plasma channels.

The remainder of this paper is organized as follows. Section II introduces the asymmetric-driver hollow-channel scheme and outlines the key physical processes involved. Section III systematically analyzes the driver evolution and identifies the criteria required for stable wake formation. Section IV summarizes the main conclusions and discusses implications for future plasma-based positron acceleration schemes.

\begin{figure*}[htb]
	\hspace*{-0.8cm}
	\includegraphics[width=1.1\linewidth]{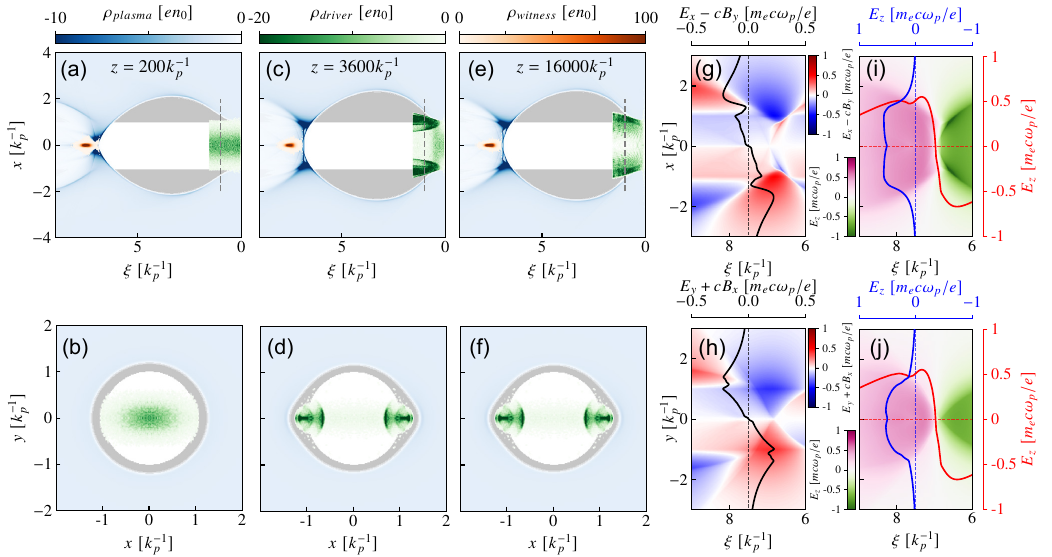}
	\caption{\label{fig:fig_1} Evolution of the driver beam, plasma wake, and positron beam in a hollow channel. 
		(a)--(b) $\xi-x$ distribution and the corresponding transverse cross section at $\xi=1\,k_p^{-1}$ (gray dash line in (a)) of plasma and beam densities at $z=200\,k_p^{-1}$.
		(c)--(d) Same at $z=3600\,k_p^{-1}$.
		(e)--(f) Same at $z=16000\,k_p^{-1}$.
		(g)--(h) $\xi-x$ and $\xi-y$ distributions of the transverse field at $z=3600\,k_p^{-1}$. 
		(i)--(j) Same for the accelerating field.}
\end{figure*}

\section{Wake Excitation and Evolution Driven by An Asymmetric Electron Beam in Hollow Plasma Channels \label{Sec:II}}
As proposed in \cite{Hollow_channel_2021_Zhou}, a transversely asymmetric electron beam propagating on axis in a hollow plasma channel excites a wakefield dominated by a quadrupole component. For an elliptical driver with, for example, $\sigma_x > \sigma_y$, the resulting wakefield provides focusing forces in the $y$ direction and defocusing forces in the $x$ direction. When the dense electron beam reaches the inner boundary of the hollow channel, plasma electrons are expelled outward while the heavier ions remain nearly stationary. The exposed ion background generates a restoring focusing force that compensates the transverse defocusing of the quadrupole wake, enabling sustained beam propagation. A key feature of this configuration is that the wakefield structure emerges self-consistently with the evolution of the driver. As the expelled plasma electrons overshoot the channel boundary and subsequently re-enter the channel, focusing forces for positrons arises, which would have been absent in a hollow plasma channel. Moreover, accelerating fields for positrons also emerge after the plasma electrons cross the axis.

The wake excitation and evolution in $\xi-x$ plane are illustrated in Fig.~\ref{fig:fig_1}(a), (c) and (e), obtained from fully 3D quasistatic PIC simulations using QuickPIC\cite{QuickPIC_2006_Huang,QuickPIC_2013_An}, where a co-moving coordinate $\xi \equiv ct-z$ is used, and $c$ is the velocity of light in vacuum. 
As shown in Fig.~\ref{fig:fig_1} (a), the plasma channel density profile is defined as $n_p (r)=n_0 [H(r-r_{\rm in} )-H(r-r_{\rm out})]$, where $H$ is the Heaviside step function, with inner and outer radii $r_{\rm in}=1.0\,k_p^{-1}$ and $r_{\rm out}=4.5\,k_p^{-1}$, respectively, where $\,k_p^{-1}$ is the skin depth of the plasma. 
The asymmetric electron driver has a flat-top longitudinal current profile, transverse sizes $\sigma_{x0}=0.4\,k_p^{-1}$, $\sigma_{y0}=0.2\,k_p^{-1}$, and a length of $L_d=1.5\,k_p^{-1}$. 
The peak driver density is $12n_0$, with an initial energy of 10 GeV and normalized emittance $\epsilon_{x0} = 0.4\,k_p^{-1}$ and $\epsilon_{y0} = 0.2\,k_p^{-1}$. 
A tri-Gaussian positron beam is loaded into wake, with a size of $\sigma_{r,e^+}=0.05\,k_p^{-1}$, $\sigma_{z,e^+}=0.15\,k_p^{-1}$ and a peak density of $n_{peak,e^+}=150n_0$. The simulation box has a range of $10\times10\times9\,k_p^{-3}$, and the spatial resolution are $0.0195\times0.0195\times0.0176\,k_p^{-3}$ in $x$, $y$ and $z$ directions, respectively. 
The box moves in the $+z$ direction at a speed of $c$. 
Throughout this paper, unless otherwise specified, all physical quantities are presented in normalized units: lengths are normalized to $\,k_p^{-1}$, densities to the ambient plasma density $n_0$, and fields to the cold waving-breaking limit $E_0=m_{e}c\omega_p/e$, where $m_e$ is the electron mass, $e$ is the elementary charge, and $\omega_p=ck_p$ is the plasma frequency. When converting to physical units we assume the plasma density is $n_p=5\times10^{16}\,{\rm cm^{-3}}$, corresponding to $\,k_p^{-1}=23.67\, {\rm \mu m}$.

As the driver propagates, the quadrupole wakefield induces a gradual transverse deformation of the beam, leading to a characteristic splitting and redistribution of the driver density, as shown in Fig.~\ref{fig:fig_1}(b), (d) and (f). This self-consistent evolution modifies the plasma electron response and progressively reshapes the wake structure. By $z=3600\,k_p^{-1}$ (Fig.~\ref{fig:fig_1}(c)--(d)), the system reaches a quasisteady state in which the wakefield profile and driver distribution evolve only weakly with further propagation. This state is maintained over distances exceeding $10^4 \,k_p^{-1}$ (Fig.~\ref{fig:fig_1}(e)--(f)), indicating long-term dynamical stability. 

The $\xi-x$ and $\xi-y$ distributions of transverse and accelerating fields at $z=3600\,k_p^{-1}$ are presented in Fig.~\ref{fig:fig_1}(g)--(j), together with the transverse lineouts taken at the positron beam loading phase. Within the hollow channel, the accelerating field exhibits a relatively flat transverse profile (Fig.~\ref{fig:fig_1}(i)--(j)), reflecting the characteristics of hollow plasma channel. However, anisotropy exists: the focusing strength in the $x$ direction is systematically weaker than that in $y$ (Fig.~\ref{fig:fig_1}(g)--(h)), consistent with the asymmetric transverse charge distribution of the driver.

It should be noted that a step-like channel wall density profile is adopted as a physically reasonable approximation, as the driver dynamics are primarily determined by the inner radius and wall density rather than the exact boundary steepness. A flat-top longitudinal current profile is chosen to provide a well-defined beam head for the theoretical analysis. The current profile redistributes the wakefield amplitude along $\xi$ but does not alter the transverse quadrupole structure of the field, which is determined by the driver's asymmetric cross-section and the channel geometry. Additional simulations confirm that the results remain valid for non-ideal channel boundaries (e.g., super-Gaussian profiles) and non-flat-top beam distributions (e.g., Gaussian profiles), as detailed in Appendix~\ref{Appendix:A}.

The existence of such a quasisteady wake is essential for sustained plasma-based acceleration. However, this regime is not universally obtained. The transition to, and stability of, the quasisteady state depends sensitively on the initial driver parameters and interaction with the channel boundary. In the following section, we systematically investigate the conditions under which the asymmetric driver evolves into a stable quasisteady state, and identify the parameter regimes that enable long-lived wakefield structures suitable for further beam–plasma interaction studies.

\section{Evolution Regimes of Asymmetric-Driver-Driven Wakes \label{Sec:III}}
While the quasisteady wake described in Sec.~\ref{Sec:II} enables simultaneous focusing and acceleration of positrons, such a wake does not arise for arbitrary transversely asymmetric drivers. In practice, the long-term evolution of the wake is determined by the initial transverse parameters of the electron driver and the geometry of the channel. If these parameters are not properly chosen, the wake structure can undergo qualitative changes that harms the foundation required by high-quality positron acceleration,

In this section, we identify the driver evolution scenarios that prevent the formation of a positron-favorable wake. Two distinct classes of undesired wake evolution are observed in PIC simulations, both of which destabilize the wake structure and preclude high-quality positron acceleration.
\subsection{Quadrupole Polarity Reversal Induced by Transverse Phase-Space Mismatch \label{Sec:IIIA}}
\begin{figure}[htb]
	\hspace*{-0.7cm}
	\includegraphics[width=1.1\linewidth]{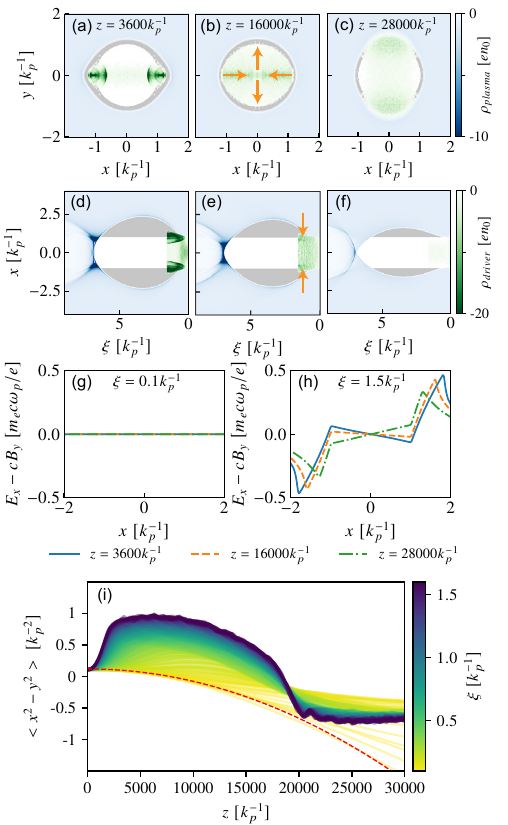}
	\caption{\label{fig:fig_2} Polarity reversal of the quadrupole field. Transverse cross section at $\xi=1\,k_p^{-1}$ at 
		(a) $z=3600\,k_p^{-1}$, 
		(b) $16000\,k_p^{-1}$ and 
		(c) $28000\,k_p^{-1}$. 
		(d)--(f) Corresponding $\xi-x$ distributions of plasma and beam densities. 
        (g) $E_x-cB_y$ experienced by the beam head slice at $\xi=0.1k_p^{-1}$.
        (h) $E_x-cB_y$ experienced by the beam tail slice at $\xi=1.5k_p^{-1}$.
		(i) Evolution of $\left\langle x^2-y^2 \right\rangle$ values for each longitudinal slice of the driver beam. The red curve is the $\sigma_x^2 (z)-\sigma_y^2 (z)$ under free drift.}
\end{figure}

Figure~\ref{fig:fig_2} illustrates the first scenario, which occurs when the driver has an unmatched initial emittance with respect to its transverse size, for example, $\epsilon_{x0}=0.2\,k_p^{-1}$  and $\epsilon_{y0}=0.2\,k_p^{-1}$, while all other parameters, including the profile of the beam and channel geometry, are identical to those in Fig.~\ref{fig:fig_1}. At an early propagation distance (Fig.~\ref{fig:fig_2} (a) and (d), $z=3600\,k_p^{-1}$), the driver splits in the $x$ direction and reaches the inner boundary of the plasma wall, exciting wake closely resembles the quasisteady structure shown in Fig.~\ref{fig:fig_1}(c). However, as the propagation continues, the quadrupole wakefield undergoes a polarity reversal, in which the beam sections merge in the $x$ direction and split in y (Fig.~\ref{fig:fig_2}(b), $z=16000\,k_p^{-1}$), and the wake structure evolves correspondingly (Fig.~\ref{fig:fig_2}(e)). Finally, the wake reaches a new quasistatic regime, with the driver staying around the boundary in $y$ (Fig.~\ref{fig:fig_2}(c), $z=28000\,k_p^{-1}$) and wake structure interchanging in $x$ and $y$ directions (Fig.~\ref{fig:fig_2}(f)). This transition signals a fundamental change in the driver–plasma interaction and alters the transverse focusing structure. 

The origin of the quadrupole polarity reversal can be traced to the transverse evolution of the driver head, which gradually erases and eventually reverses its initial transverse asymmetry during propagation. As the downstream wake is determined by the upstream charge distribution, the evolution of the head plays a decisive role in setting the quadrupole component of the wakefield. Since the beam head experiences negligible wakefield (Fig.~\ref{fig:fig_2}(g)), its transverse dynamics can be approximated as free drift. In contrast, the trailing slices do not follow free-drift dynamics; instead, they experience a significant and evolving transverse wakefield generated by the upstream slices (Fig.~\ref{fig:fig_2}(h)). The strong transverse forces and subsequent redistribution of the beam tail observed in simulations are the self-consistent consequences of this head-driven wakefield inversion. The quadrupole component of the transverse field can be expressed as \cite{Hollow_channel_1999_Schroeder,Hollow_channel_2021_Zhou} 
\begin{equation}
W_{\bot2}\propto \left\langle x^2-y^2 \right\rangle (x\hat{x}-y\hat{y})+\left\langle 2xy\right\rangle (y\hat{x}+x\hat{y}),
\end{equation}
where $\left\langle x^2-y^2 \right\rangle\approx\sigma_x^2-\sigma_y^2$, and the second term can be neglected in the present configuration. Under free drift, the transverse size of the beam head in the $x$ direction evolves as 
\begin{equation}
	\sigma_x (z)=\sqrt{\sigma_{x0}^2+\frac{\epsilon_{x0}^2}{\gamma^2 \sigma_{x0}^2}z^2},
\end{equation}
and the analogous equation for $\sigma_y$ is obtained with $\sigma_{x0}\rightarrow\sigma_{y0}$, $\epsilon_{x0}\rightarrow\epsilon_{y0}$, where $\gamma$ is the Lorentz factor of beam head (approximately constant).
For sufficiently long propagation distances, this reduces to $\sigma_x (z)\approx (\epsilon_{x0}/\gamma\sigma_{x0})z$, $\sigma_y (z)\approx (\epsilon_{y0}/\gamma\sigma_{y0})z$. Consequently, for $\epsilon_{x0}=0.2\,k_p^{-1}, \epsilon_{y0}=0.2\,k_p^{-1}$ and $\sigma_{x0}=0.4\,k_p^{-1}, \sigma_{y0}=0.2\,k_p^{-1}$, $\sigma_x (z)$ will eventually be smaller than $\sigma_y (z)$, causing  $\left\langle x^2-y^2 \right\rangle$ to change sign and leading to a reversal of the quadrupole polarity. 

Figure~\ref{fig:fig_2}(i) presents evolution of $\left\langle x^2-y^2 \right\rangle$ values of each slice, confirming the process. The red dash curve is the $\sigma_x^2 (z)-\sigma_y^2 (z)$ value of beam head calculated analytically, which shows a good agreement with the PIC simulation result. 
Except for the leading slice, the downstream slices deviate from the red dashed curve in $\left\langle x^2-y^2 \right\rangle$ because they have already experienced significant transverse wakefield forces and therefore no longer follow the free-drift evolution (Fig.~\ref{fig:fig_2}(h)). Nevertheless, slices closer to the head exhibit earlier sign reversal of $\left\langle x^2-y^2 \right\rangle$, confirming that the quadrupole polarity reversal propagates from the head toward the trail. 

Since the focusing strength of positrons in this scheme differs in transverse directions and relies on the polarity of the quadrupole wake as shown in Fig.~\ref{fig:fig_1}(g) and (h), a reversal of the quadrupole field fundamentally alters the focusing environment, which can lead to an increase of the positron emittance and even loss of charge. To preserve the quadrupole polarity, and hence a positron-favorable focusing structure, one needs
\begin{equation}
\frac{\epsilon_{x0}}{\sigma_{x0}} \geq\frac{\epsilon_{y0}}{\sigma_{y0}} ,\sigma_{x0}>\sigma_{y0} 
\quad \text{or}
\quad \frac{\epsilon_{x0}}{\sigma_{x0}} \leq\frac{\epsilon_{y0}}{\sigma_{y0}} ,\sigma_{x0}<\sigma_{y0},
\label{eq:polarity_condition}
\end{equation}
which ensures that the initial transverse asymmetry of the driver is maintained throughout propagation.

\subsection{Continuous Penetration into the Plasma Wall Due to Excessive Transverse Kinetic Energy \label{Sec:IIIB}}
\begin{figure}[htb]
	\hspace*{-0.5cm}
	\includegraphics[width=1.1\linewidth]{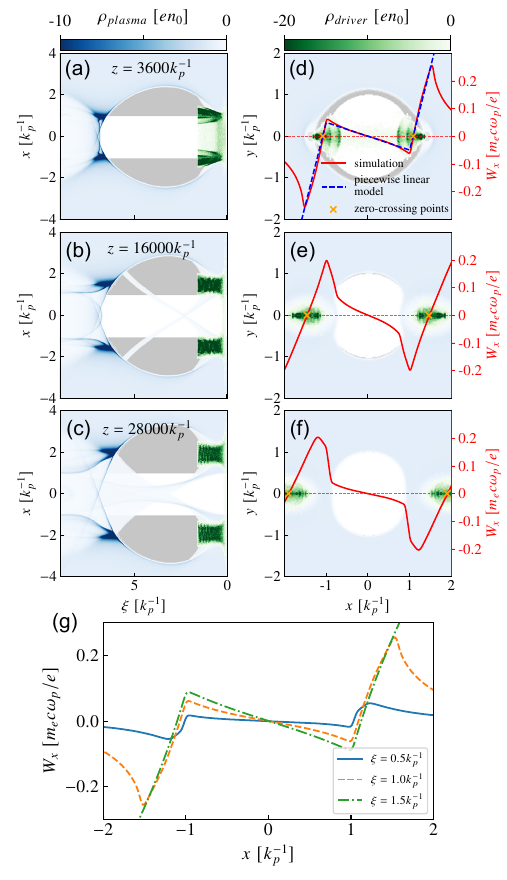}
	\caption{\label{fig:fig_3} Continuous penetration of the electron driver. $\xi-x$ distributions of plasma and beam densities at 
		(a) $z=3600\,k_p^{-1}$, 
		(b) $16000\,k_p^{-1}$, 
		(c) $28000\,k_p^{-1}$. 
		(d)--(f) Snapshots of the transverse cross section at $\xi=1\,k_p^{-1}$; the red line is the transverse field $W_x$ at $y=0$. 
        (g) Lineouts of $W_x$ at $\xi=0.5\,k_p^{-1}$, $1.0\,k_p^{-1}$ and $1.5\,k_p^{-1}$ at $z=3600\,k_p^{-1}$.
    }
\end{figure}

In contrast to the polarity reversal caused by mismatch of the initial emittance with respect to the transverse size, an excessively large transverse kinetic energy destabilizes the wake through a qualitatively different mechanism. As shown in Fig.~\ref{fig:fig_3}(a-c), when $\epsilon_{x0}=1.2\,k_p^{-1}$ and $\epsilon_{y0}=0.2\,k_p^{-1}$, the separated driver sections in the $x$ direction no longer undergo bounded oscillations near the inner channel boundary. Instead, they continuously penetrate into the plasma wall, preventing the wake from evolving into a quasisteady configuration and disrupting the self-consistent electron return into the hollow region. 

Figures~\ref{fig:fig_3}(d-f) present the transverse density distribution of the driver and plasma and the corresponding transverse field ($W_x \equiv E_x-cB_y$) lineouts at $\xi=1\,k_p^{-1}$ and $y=0$. As shown in Fig.~\ref{fig:fig_3}(d), inside the channel, the quadrupole field provides a net defocusing field in the $x$ direction. As the driver expels the plasma electrons, an ion-rich region is formed near the inner channel boundary, generating an attractive transverse force that counteracts the quadrupole defocusing. As a result, $W_x$ crosses zero within this exposed ion region, defining an effective equilibrium position for the driver particles. 

If the transverse kinetic energy of the driver is limited, most particles oscillate around this zero-crossing points, and the exposed ion region remains localized near the channel boundary. In this case, the zero-crossing position of $W_x$ is nearly stationary, allowing the wake to reach the quasisteady regime discussed in Sec.~\ref{Sec:II}. However, when the driver possesses sufficient transverse kinetic energy, a substantial fraction of particles penetrate deeply into the plasma wall, as shown in Fig.~\ref{fig:fig_3}(e) and (f). This deeper penetration modifies the spatial extent of the exposed ion region, causing the zero-crossing points of $W_x$ to shift together with the driver centroid. As a result, the transverse focusing structure no longer acts as a restoring force but instead follows the driver motion. This loss of a fixed equilibrium position leads to irreversible penetration of the driver into the plasma wall and prevents the wake from evolving into a stationary state. In this wall-penetration regime, the focusing and accelerating fields exhibit persistent temporal and spatial variations. Such a non-stationary wake is expected to be unfavorable for maintaining high-quality positron acceleration.

A fully quantitative criterion for avoiding the wall-penetration regime would require a self-consistent description of the nonlinear plasma response, which varies strongly along $\xi$ and evolves with the driver. Such a description is beyond the scope of the present work. Instead, we construct a reduced model at the instant when the driver reaches the inner wall boundary to capture the essential physical mechanism responsible for the transition to the wall-penetration regime. We consider a normalized piecewise-linear model $\tilde{W_x}$ to approximate the simulated profile in Fig.~\ref{fig:fig_3}(d). The transverse coordinate and channel inner radius are normalized as $\tilde{x}=k_p x$, $\tilde{r_{\rm in}}=k_p r_{\rm in}$. The model is expressed as:
\begin{equation}
	\tilde{W}_x(\tilde{x}) =
	\begin{cases}
		k_1 \tilde{x}, & |\tilde{x}| \leq \tilde{r}_{\rm in}, \\
		k_2 \tilde{x} + (k_1-k_2)\tilde{r}_{\rm in}, & |\tilde{x}| > \tilde{r}_{\rm in}
	\end{cases}
\label{eq:piecewise_linear_model}
\end{equation}
where $k_1$ and $k_2$ are the dimensionless field gradients normalized to $E_0 k_p$. In the internal region ($|\tilde{x}| \leq \tilde{r}_{\rm in}$), the negative gradient $k_1<0$ characterizes the defocusing quadrupole field, as for an electron beam ($q=-e$), the transverse force $F_x=-eW_x$ yields $\partial{F_x}/\partial{x}=-eE_0 k_p k_1>0$. Conversely, in the plasma wall ($|\tilde{x}| > \tilde{r}_{\rm in}$), the positive gradient $k_2>0$ provides a restoring force that directs the electrons back toward the axis. The intercept $(k_1-k_2)\tilde{r}_{\rm in}$ ensures the continuity of the transverse field at the vacuum-plasma interface $\tilde{r}_{\rm in}$. At $\xi=1\,k_p^{-1}$, the values $k_1=-0.05$ and $k_2=0.6$ provide a reasonable fit to the simulation results, as shown in Fig.~\ref{fig:fig_3}(d). 

Within this simplified picture, driver particles penetrating into the plasma wall experience an effective potential barrier determined by the transverse wakefield. If particles possess a transverse kinetic energy that enable them to penetrate into the plasma wall by a depth exceeding $\sigma_{x0}$, over 50\% of the particles will enter the plasma wall (assuming Gaussian-like distribution). In this situation, the spatial extent of the exposed ion region is modified by the driver itself, causing the equilibrium position (zero-crossing points of $W_x$) to shift together with the driver. This feedback prevents the establishment of a quasisteady restoring structure and leads to continuous penetration. To estimate when this transition can occur, we compare the average transverse kinetic energy per particle with the potential barrier height $\Delta U$. The average transverse kinetic energy is given by: 
\begin{equation}
	\left\langle E_{k,x} \right\rangle=\frac{1}{2} \gamma m_e \left\langle v_x^2 \right\rangle =\frac{1}{2\gamma } \frac{\epsilon_{x0}^2}{\sigma_{x0}^2} m_{e}c^2.
\end{equation}
This kinetic energy must be overcome by the potential change $\Delta U$ provided by $W_x$ between the axis ($\tilde{x}=0$) and $\tilde{x}=\tilde{r}_{\rm in}+\tilde{\sigma}_{x0}$, with $\tilde{\sigma}_{x0}=k_p \sigma_{x0}$:
\begin{equation}
	\begin{aligned}
		\Delta U
		&= m_{e}c^2 \int_0^{\tilde{r}_{\rm in}+\tilde{\sigma}_{x0}}
		\tilde{W}_x(\tilde{x})\, d\tilde{x} \\
		&= m_{e}c^2 \left(
		\int_0^{\tilde{r}_{\rm in}} k_1 \tilde{x}\, d\tilde{x}
		+
		\int_0^{\tilde{\sigma}_{x0}} (k_2 u + k_1 \tilde{r}_{\rm in})\, du
		\right) \\
		&= \left(
		\frac{1}{2} k_1 \tilde{r}_{\rm in}^2
		+ \frac{1}{2} k_2 \tilde{\sigma}_{x0}^2
		+ k_1 \tilde{r}_{\rm in} \tilde{\sigma}_{x0}
		\right) m_{e}c^2.
	\end{aligned}
\end{equation}

A necessary condition for beam confinement is that the potential barrier is sufficient to contain the transverse motion, which leads to:
\begin{equation}
	\frac{1}{2} k_1 \tilde{r}_{\rm in}^2
	+ \frac{1}{2} k_2 \tilde{\sigma}_{x0}^2
	+ k_1 \tilde{r}_{\rm in} \tilde{\sigma}_{x0}
	>
	\frac{1}{2\gamma}\frac{\epsilon_{x0}^2}{\sigma_{x0}^2}.
	\label{eq:confinement_condition}
\end{equation}

It is important to note that the transverse force coefficients $k_1$ and $k_2$ vary significantly along the driver due to the transient buildup of the plasma response (Fig.~\ref{fig:fig_3}(g)). Consequently, the effective barrier $\Delta U$ can differ by orders of magnitude between longitudinal slices. While reducing the driver length can suppress this head-tail variation in the linear regime \cite{Hollow_channel_1999_Schroeder}, this does not eliminate the variation, and a representative slice is adopted below. The slice evaluated here (at $\xi=1\,k_p^{-1}$) is selected because the transverse restoring force has built up to a sufficiently high magnitude, allowing it to capture the characteristic confinement fields experienced by the main body of the driver. Therefore, the above condition does not define a sharp, universal stability boundary for the entire beam, but rather provides qualitative insight into the dominant trends. In particular, $\Delta U$ has no explicit dependence on the initial emittance $\epsilon_{x0}$, implying that smaller emittance reduces the likelihood that particles overcome the effective barrier. Moreover, for fixed field strengths, reducing the inner channel radius $r_{\rm in}$ increases $\Delta U$, thereby enhancing transverse confinement.

\begin{figure}[htb]
	\centering
	\includegraphics[width=1.0\linewidth]{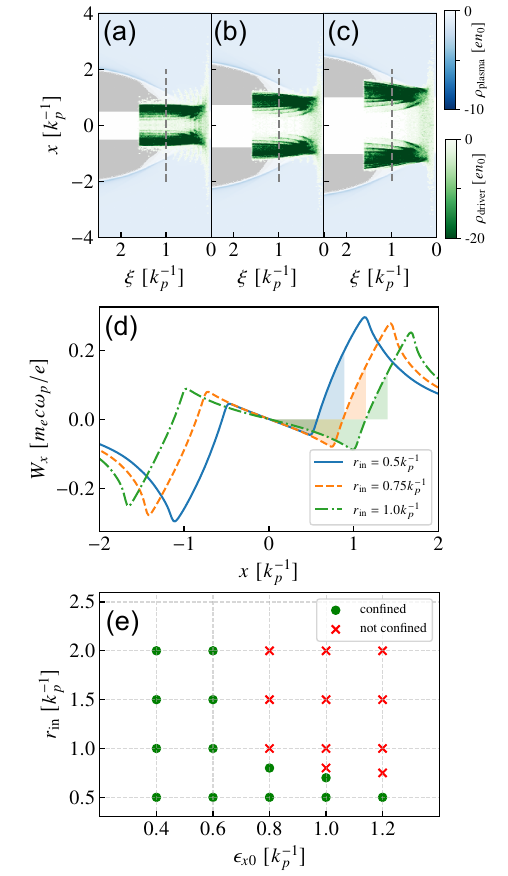}
	\caption{\label{fig:fig_4}
		Driver--plasma distributions for three inner radii when the driver reaches the boundary: 
		(a) $r_{\rm in} = 0.5\,k_p^{-1}$, 
		(b) $r_{\rm in} = 0.75\,k_p^{-1}$, and 
		(c) $r_{\rm in} = 1.0\,k_p^{-1}$. 
		(d) Lineouts of $W_x$ at $y = 0$ and $\xi = 1\,k_p^{-1}$. 
		(e) Parameter scan in the $\epsilon_{x0}$--$r_{\rm in}$ space. 
		The confined cases (green circles) correspond to quasisteady wakefields, while the penetration cases (red crosses) indicate continuous crossing of the plasma wall.}
\end{figure}

Simulation results support this picture. Figure~\ref{fig:fig_4}(a--c) compares the drivers with identical initial parameters propagating in channels with $r_{\rm in} = 0.5\,k_p^{-1}$, $0.75\,k_p^{-1}$, and $1.0\,k_p^{-1}$. The corresponding lineouts of $W_x$ at $\xi = 1\,k_p^{-1}$ are presented in Fig.~\ref{fig:fig_4}(d). The net positive area of the $W_x$ profile, defined as the shaded area above $W_x=0$ minus that below $W_x=0$, represents $\Delta U$, which clearly increases as $r_{\rm in}$ is reduced. Figure~\ref{fig:fig_4}(e) illustrates the results of a parametric scan over the initial $\epsilon_{x0}$ and $r_{\rm in}$ ($\epsilon_{y0} = 0.2\,k_p^{-1}$ is maintained) to determine the boundary for beam confinement. For a high-emittance driver with $\epsilon_{x0} = 1.2\,k_p^{-1}$, stable confinement is only maintained when the channel is sufficiently narrow, specifically for $r_{\rm in} \lesssim 0.5\,k_p^{-1}$. For $\epsilon_{x0} = 1.0\,k_p^{-1}$, the beam remains confined at $r_{\rm in} = 0.7\,k_p^{-1}$, but penetrates into the plasma at larger $r_{\rm in}$. As the initial emittance decreases to $\epsilon_{x0} = 0.8\,k_p^{-1}$, the confinement threshold for the channel radius increases to $r_{\rm in} \lesssim 0.8\,k_p^{-1}$. Notably, for drivers with $\epsilon_{x0} \leq 0.6\,k_p^{-1}$, the beam remains stably confined near the inner plasma wall even for a significantly wider channel of $r_{\rm in} = 2.0\,k_p^{-1}$. Overall, these PIC simulation results are physically consistent with our analytical model, confirming that a smaller initial emittance and a narrower channel radius facilitate stable beam confinement.

\subsection{Parameter Regimes Supporting Quasi-Steady Wakes \label{Sec:IIIC}}
Taken together, the above results show that the transverse stability of an asymmetric electron driver in a hollow plasma channel is limited by two distinct mismatch mechanisms. When the initial emittance is incompatible with the imposed beam asymmetry, the head of the driver reverses its initial transverse asymmetry during free drift, leading to a reversal of the quadrupole wake polarity. In contrast, when the transverse kinetic energy is too large relative to the confinement provided by the ions, a significant fraction of driver particles can penetrate through the plasma wall, preventing the wake from reaching a quasisteady configuration.

These two constraints together define a finite parameter window in which a long-lived, quadrupole-dominated wake can be sustained. 
Preserving the quadrupole polarity requires the initially wider beam dimension to be maintained [Eq.~(\ref{eq:polarity_condition})].
Avoiding continuous wall penetration further requires that the transverse potential well be sufficiently deep to confine the driver particles [Eq.~(\ref{eq:confinement_condition})], which should be regarded as a necessary condition.
In this work, 
\begin{equation}
	\frac{\epsilon_{x0}}{\sigma_{x0}} \simeq \frac{\epsilon_{y0}}{\sigma_{y0}} \simeq 2,
	\qquad
	r_{\rm in} \simeq 2\sigma_{x0} \simeq 4\sigma_{y0},
\end{equation}
serve as a representative choice.

\subsection{Positron Beam Transport in Different Wake Evolution Regimes \label{Sec:IIID}}
\begin{figure}[htb]
%	\hspace*{-0.8cm}
	\includegraphics[width=1.0\linewidth]{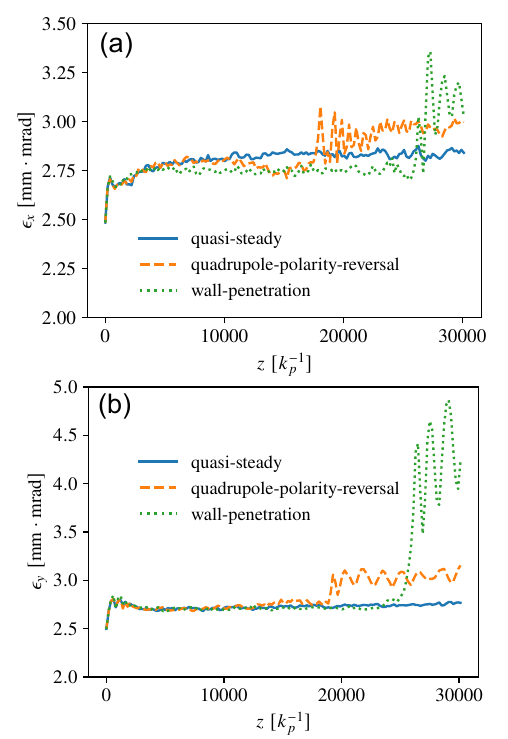}
	\caption{\label{fig:fig_5}
		 Evolution of projected emittance of the positron beams (a) in $x$ direction and (b) in $y$ direction. 
		The initial hollow plasma channel \((r_{\rm in}=1.0\,k_p^{-1},\, r_{\rm out}=4.5\,k_p^{-1})\), positron beam \((\sigma_{r,e^+}=0.05\,k_p^{-1},\, \sigma_{z,e^+}=0.15\,k_p^{-1},\, n_{\rm peak,e^+}=150\,n_0,\, E_0=1\,{\rm GeV},\, \epsilon_{n,e^+}=1.0\,k_p^{-1})\), and electron driver size and density \((\sigma_{x0}=0.4\,k_p^{-1},\, \sigma_{y0}=0.2\,k_p^{-1},\, L_d=1.5\,k_p^{-1},\, n_{\rm peak,d}=12\,n_0)\) are fixed in all cases. Only the initial driver emittances $(\epsilon_{x0}, \epsilon_{y0})$ differ: quasisteady \((0.4,0.2)\,k_p^{-1}\), quadrupole-polarity-reversal \((0.2,0.2)\,k_p^{-1}\), and wall-penetration \((1.2,0.2)\,k_p^{-1}\).
		}
\end{figure}

To assess the impact of different wake evolution regimes on positron beam dynamics, we load an identical tri-Gaussian positron beam into wakes driven by the electron beams discussed in Sec.~\ref{Sec:II}, \ref{Sec:IIIA} and \ref{Sec:IIIB}.
Here, the trailing positron bunch serves primarily as an unoptimized diagnostic test beam to probe the wakefield structures, with its specific parameters summarized in the caption of Fig.~\ref{fig:fig_5}. To match the short plasma wavelength while maintaining a total charge of $\sim$100 pC, a compact bunch is adopted, which naturally evaluates to a peak density of $150\,n_0$.
This setup allows for a controlled comparison of positron beam response under different wake evolution regimes.

Figure~\ref{fig:fig_5} compares the evolution of the projected emittance of the positron beams in the transverse directions for the three cases. In the quasisteady regime (blue solid curves), the projected emittance exhibits a rapid initial increase by about 12\% followed by clear saturation over the remaining propagation distance. This behavior indicates that after an initial transient, the positron beam experiences a time-independent focusing environment, consistent with the formation of a long-lived, quasisteady wake, and is observed previously \cite{Hollow_channel_2021_Zhou}.
In practical experimental implementations, a working gas with a high ionization potential, such as helium, can be employed to prevent the potential beam-induced field ionization of residual neutral gas within the channel core. Self-consistent 3D PIC simulations incorporating field ionization\cite{Ionization_2025_Tang}, as detailed in Appendix~\ref{Appendix:B}, confirm that no beam-induced helium ionization occurs under our parameters, resulting in no impact on wake structure.

In the quadrupole polarity reversal regime (orange dashed curves), the projected emittance also saturates at early times, but undergoes a second distinct growth phase that is temporally correlated with the reversal of the quadrupole wake polarity ($z\approx20000\,k_p^{-1}$). This additional emittance growth reflects the response of the positron beam to the reorientation of the transverse focusing axes, which induces renewed phase mixing. After the reversal, the emittance again approaches saturation once the wake relaxes into a new quasisteady configuration. The final emittance increase is approximately a factor of two larger than in the quasisteady case.

In contrast, in wall-penetration regime (green dotted curves), the wake fails to evolve into a stationary state, leading to a significant and continuous growth of the positron emittance over long distances. Initially, as the wakefield deviates only slightly from the quasisteady structure, the positron beam maintains a relatively stable focusing environment by attracting plasma electrons toward the axis\cite{Hollow_channel_2021_Zhou}, resulting in a temporary saturation of the emittance. However, as the driver continues its outward movement, the distribution of plasma electrons entering the channel undergoes drastic changes. the original focusing structure experienced by positrons eventually collapses, causing the projected emittance to grow significantly without any indication of saturation.

These results demonstrate that the evolution regime of the asymmetric-driver-driven wake directly determines whether a stable environment for high-quality positron acceleration can be established. Detailed optimization of beam loading and positron beam matching is beyond the scope of this work and will be addressed in a separate publication.

\section{Discussion and Conclusion \label{Sec:IV}}
In this work, we investigated the evolution regimes of wakefield driven by a transversely asymmetric electron beam in a hollow plasma channel using fully 3D PIC simulations. 
Two distinct mechanisms that prevent the formation of a stable wake are identified. When the initial emittance of the driver is incompatible with the imposed beam asymmetry, the free drift of the beam head progressively erases the initial transverse asymmetry, leading to a reversal of the quadrupole wake polarity. Conversely, when the driver possesses a large transverse kinetic energy, driver particles can overcome the effective confining potential near the channel boundary, resulting in continuous penetration into the plasma wall and the absence of a steady wake. Although these two scenarios arise from different physical processes, both disrupt the self-consistent electron dynamics required to sustain a usable accelerating structure. 
Based on these observations, practical parameter regimes of quasisteady wakes are proposed, which links the driver emittance, transverse size, and channel geometry. While not representing a sharp boundary in parameter space, these regimes provide a practical guideline for selecting driver parameters that favor long-term wake stability and is readily applicable in numerical and experimental designs. A systematic mapping of the stability boundary in parameter space is left for future work.

We further examine positron beam transport in different wake evolution regimes. Even without positron beam optimization, clear qualitative differences are observed. In the quasisteady regime, the positron beam experiences a stable focusing and accelerating environment, leading to bounded emittance growth. In contrast, unstable wake evolution introduces time-dependent transverse fields that introduce extra emittance growth. These results demonstrate that wake stability is a necessary prerequisite for high-quality positron acceleration.

The present study therefore lays the groundwork for subsequent investigations focused on positron beam optimization, which can be meaningfully pursued once wake stability is ensured. 
The optimization of positron beam quality in this scheme, including beam loading and emittance matching, remains a topic for future study.

\begin{acknowledgments}
We thank Professor Weiming An from Beijing Normal University for his useful discussions. This work was supported by the Strategic Priority Research Program of the Chinese Academy of Sciences (XDB0530000), the National Natural Science Foundation of China (NSFC) (Grants No. 12305152, No. 12375241, No. 12405169), Research Foundation of Institute of High Energy Physics, Chinese Academy of Sciences (Grants No. E65152U1, No. E65459U8), Ministry of Science and Technology of the People’s Republic of China (Grant No. 2023YFA1606300), Science Fund Program for Distinguished Young Scholars (Overseas), the Discipline Construction Foundation of “Double World-class Project”. The simulation work is supported by Center of High Performance Computing, Tsinghua University.
\end{acknowledgments}

\appendix
\counterwithin{figure}{section}
\counterwithin{table}{section}
\counterwithin{equation}{section}

\renewcommand{\thefigure}{\thesection\arabic{figure}}
\renewcommand{\thetable}{\thesection\arabic{table}}
\renewcommand{\theequation}{\thesection\arabic{equation}}

\setcounter{figure}{0}
\setcounter{table}{0}
\setcounter{equation}{0}

\section{Robustness of Driver Evolution against Nonideal Channel Boundaries and Beam Profiles \label{Appendix:A}}

In practical experiment, hollow plasma channels are typically generated via optical field ionization of a neutral gas using structured laser pulses, such as high-order Bessel beams. This process naturally results in a smoothed boundary profile.

\begin{figure}[htb]
    \hspace*{-0.5cm}
    \includegraphics[width=1.0\linewidth]{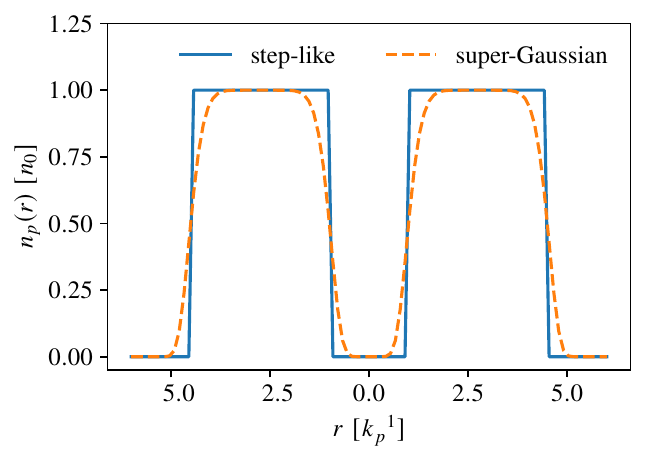}
    \caption{
    Comparison of the step-like and super-Gaussian plasma channel profiles.}
    \label{fig:app:A1} 
\end{figure}

\begin{figure}[htb]
    \hspace*{-0.5cm}
    \includegraphics[width=1.0\linewidth]{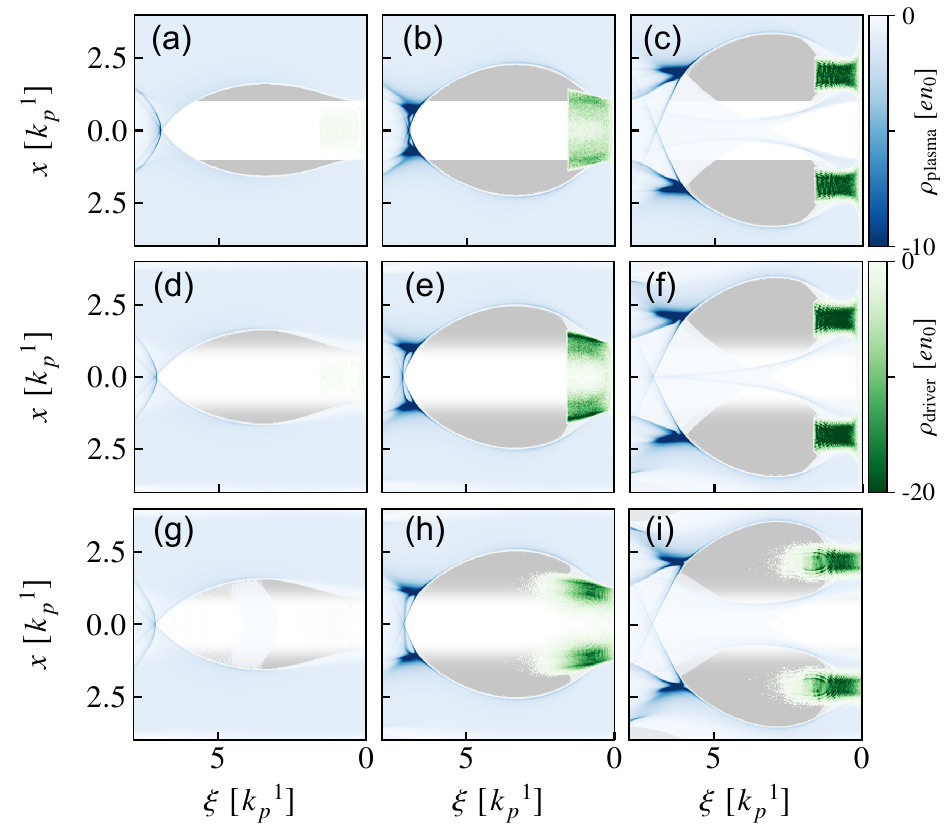}
    \caption{
    Comparison of the driver evolution with different longitudinal profiles in step-like and super-Gaussian channels at $z=28000k_p^{-1}$. (a)-(c) Flat-top longitudinal profile in a step-like channel for quadrupole-polarity-reversal, quasisteady, and wall-penetration regimes. (d)-(f) Flat-top longitudinal profile in a super-Gaussian channel for the same regimes. (g)-(i) Gaussian longitudinal profile in a super-Gaussian channel for the same regimes.}
    \label{fig:app:A2} 
\end{figure}

To evaluate the influence of this non-step-like boundary, we performed simulations using a super-Gaussian channel wall profile, defined as 
\begin{equation}
    n(r) = n_0 \exp\left[-\frac{(r-r_0)^n}{2\sigma_0^n}\right],
\end{equation}
with $r_0=2.75k_p^{-1}$, $\sigma_0=1.68k_p^{-1}$ and $n=8$ to match the inner and outer boundary dimensions used in the main text (Fig.~\ref{fig:app:A1}). Figure~\ref{fig:app:A2} shows a comparison of the driver evolution at $z=28000k_p^{-1}$ in the step-like channel (Figs.~\ref{fig:app:A2}(a)-(c)) and in the super-Gaussian channel (Fig.~\ref{fig:app:A2} (d)-(f)). Across all three evolution regimes discussed in the main text (the quadrupole-polarity-reversal regime (left column), the quasisteady regime (middle column), and the wall-penetration regime (right column)), the transverse evolution of the asymmetric driver beam and the self-consistent wakefield structures exhibit remarkable consistency with the idealized step-profile scenario. This confirms that the step-function approximation effectively captures the core beam-plasma dynamics.

Furthermore, while a flat-top longitudinal current profile is adopted in the main text, the regime classification is unaffected by the current profile, as the current profile does not alter the transverse quadrupole structure. To verify this, additional simulations have been conducted with a more realistic Gaussian longitudinal current profile. A tri-Gaussian driver beam centered at $\xi=0.85k_p^{-1}$ with $\sigma_x=0.4k_p^{-1}$, $\sigma_y=0.2k_p^{-1}$ and $\sigma_z=0.6k_p^{-1}$ and a peak density of 12.3$n_0$ is utilized to maintain the same total charge as the flat-top beam used in the main text. Moreover, a super-Gaussian channel profile identical to that in Figs.~\ref{fig:app:A2}(d)-(f) is applied. As shown in Fig.~\ref{fig:app:A2}(g)-(i), the driver beam systematically enters the three distinct evolution regimes.

\section{Effects of residual gas inside the channel \label{Appendix:B}}

In experiments, hollow channels are created by ionizing an annular region of neutral gas with high-order Bessel laser pulses. As a result, the hollow channel consists of an annular ionized region surrounding a core of neutral gas. The intense electromagnetic fields of the driver and positron beams can potentially ionize this residual gas. However, this potential ionization can be avoided by selecting a working gas with a sufficiently high ionization potential.

\begin{figure}[htb]
    \hspace*{-0.5cm}
    \includegraphics[width=1.0\linewidth]{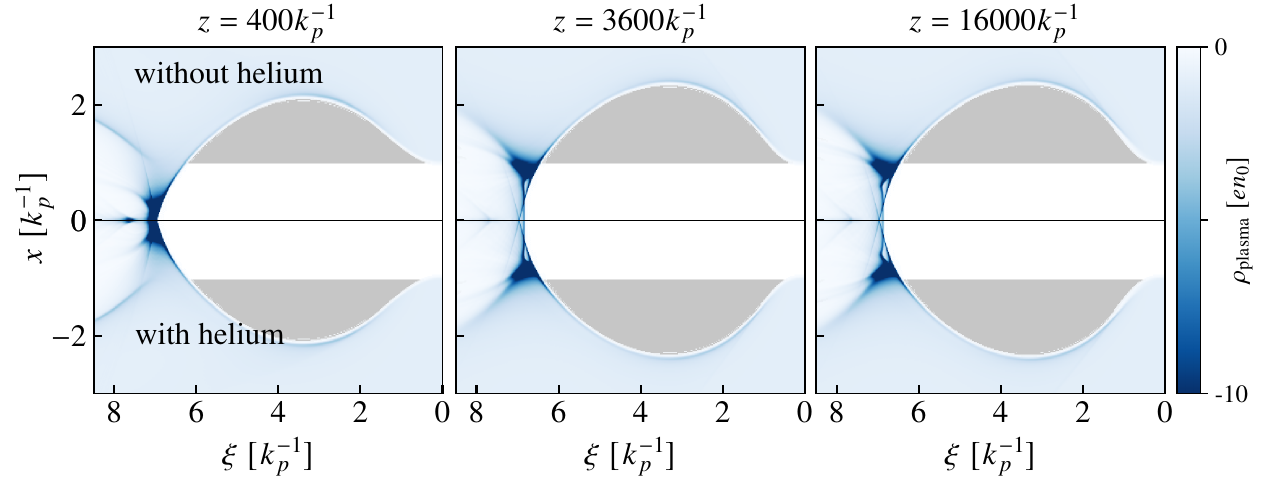}
    \caption{
      Comparison of the total electron density of the wake with and without residual gas ionization. upper row: without helium; lower row: with helium.}
    \label{fig:app:B1}
\end{figure}

For instance, employing helium prevents beam-induced ionization under the proposed beam parameters in the main text. To verify this, we performed 3D PIC simulations incorporating the ADK field-ionization model. The exact beam and plasma parameters from the quasisteady scenario in Fig.~\ref{fig:fig_5} are utilized, with the addition of a uniform background of residual neutral gas (helium with a density of 0.5$n_0$) distributed throughout the channel core. Figure~\ref{fig:app:B1} compares the total electron density of the wake with and without the residual gas ionization. The results clearly demonstrate that no beam-induced ionization occurs for the given beam parameters, and the wake maintains its structure.

\end{document}